%% file: main.tex
\documentclass{article}

\usepackage[final]{tackling_climate_workshop_style}

\input{math_commands.tex}

\usepackage{hyperref}
\usepackage{url}
\usepackage[utf8]{inputenc} 
\usepackage[T1]{fontenc}    
\usepackage{hyperref}       
\usepackage{url}            
\usepackage{booktabs}       
\usepackage{amsfonts}       
\usepackage{nicefrac}       
\usepackage{microtype}      
\usepackage{graphicx}
\usepackage{subcaption}
\usepackage{siunitx}

\title{Predicting NOx emissions in Biochar Production Plants using Machine Learning}

\author{%
    Marius Köppel\thanks{corresponding author \texttt{mkoepp\@phys.ethz.ch}} \\
    ETH Zurich
    \And
    Niklas Witzig \\
    Johannes Gutenberg University Mainz \\
    \And
    Tim Klausmann \\
    Johannes Gutenberg University Mainz \\
    \And
    Mattia Cerrato \\
    Johannes Gutenberg University Mainz \\
    \And
    Tobias Schweitzer \\
    AIRA-Holding GmbH
    \And
    Jochen Weber \\
    PYREG GmbH \\
    \And
    Erdem Yilmaz \\
    PYREG GmbH \\
    \And
    Juan Chimbo \\
    ARTi Inc
    \And
    Bernardo del Campo \\
    ARTi Inc
    \And
    Lissete Davila \\
    ARTi Inc
    \And
    David 	Barreno \\
    ARTi Inc \\
}
\begin{document}

\maketitle

\begin{abstract}
The global Biochar Industry has witnessed a surge in biochar production, with a total of 350k mt/year production in 2023.
With the pressing climate goals set and the potential of Biochar Carbon Removal (BCR) as a climate-relevant technology, scaling up the number of new plants to over 1000 facilities per year by 2030 becomes imperative.
However, such a massive scale-up presents not only technical challenges but also control and regulation issues, ensuring maximal output of plants while conforming to regulatory requirements.

In this paper, we present a novel method of optimizing the process of a biochar plant based on machine learning methods.
We show how a standard Random Forest Regressor can be used to model the states of the pyrolysis machine, the physics of which remains highly complex.
This model then serves as a surrogate of the machine -- reproducing several key outcomes of the machine -- in a numerical optimization.
This, in turn, could enable us to reduce NOx emissions -- a key regulatory goal in that industry -- while achieving maximal output still.
In a preliminary test our approach shows remarkable results, proves to be applicable on two different machines from different manufacturers, and can be implemented on standard Internet of Things (IoT) devices more generally.

\end{abstract}


\begin{figure}[h]
    \centering
    \includegraphics[width = 0.7\linewidth]{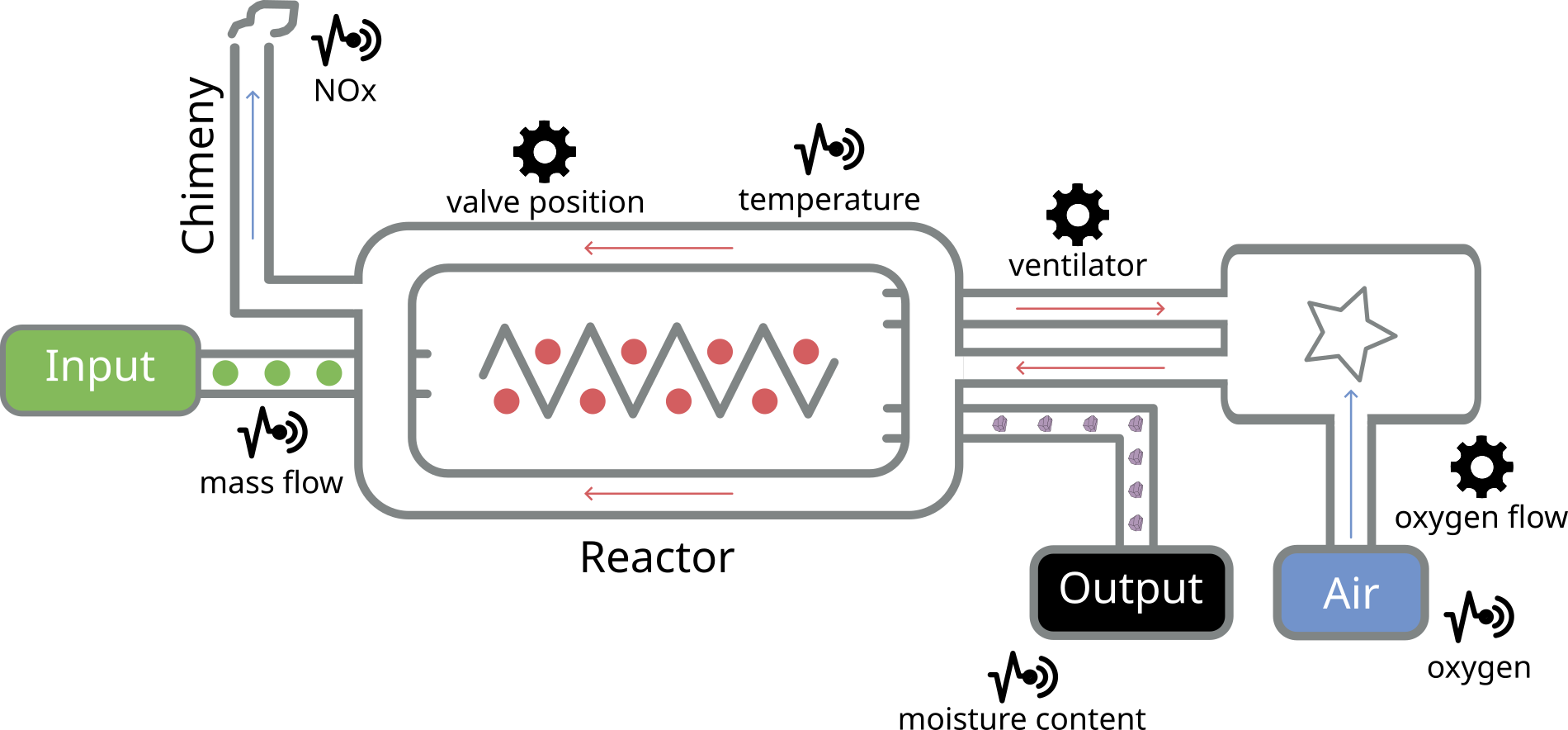}
     \caption{Sketch of a general continues pyrolysis reactor. The screws (e.g. valve position or ventilator) are the ``machine states'' (input features), while the sensor symbols (e.g. temperature or NOx) are the target which the Random Forest is predicting given the machine states. Note that the ARTi reactor has multiple champers so this image only shows a very basic design. For a more detailed representation we encounter the reader to visit the websites \citep{pyreg} and \citep{arti}.}
     \label{fig:pyreg}
\end{figure}

\section{Introduction}
The urgency of addressing climate change has led to a global call for innovative solutions that can effectively mitigate ever-increasing levels of atmospheric carbon dioxide (CO2)~\citep{negemissions}.
In line with this imperative, Carbon Dioxide Removal (CDR) strategies emerged as a crucial component of the climate action toolkit.
CDR encompasses a spectrum of anthropogenic activities aimed at the removal and durable storage of CO2 in geological, terrestrial, or oceanic reservoirs, as well as in various products~\citep{board2019negative}.
This approach holds the potential to not only suppress emissions but also to achieve net-negative CO2 concentrations in the atmosphere, essential for achieving the goal of net-zero greenhouse gas (GHG) emissions~\citep{smith2016biophysical}.
As highlighted in the Intergovernmental Panel on Climate Change (IPCC) Sixth Assessment Report (AR6) Working Group III~\citep{ipcc}, CDR strategies are indispensable, especially for counterbalancing residual emissions that are difficult to eliminate through conventional means~\citep{rogelj2018scenarios}.

The deployment of CDR solutions, however, comes with its own set of challenges, particularly with respect to feasibility and sustainability constraints, especially at larger scales.
As outlined in the IPCC AR6, upscaling CDR deployment necessitates the development of effective approaches, a task that requires innovative technologies and strategies capable of both removing CO2 from the atmosphere and ensuring its long-term storage.
Among the array of CDR methods discussed in the report, biochar stands out as a promising yet underexplored avenue.



Given the rapid growth in global biochar production from 95,000 metric tons per year in 2021 to 350,000 metric tons per year in 2023~\citep{ibi}, this trend aligns with climate action goals, positioning Biochar Carbon Removal (BCR) as a potentially crucial climate-relevant technology.
To meet ambitious climate targets, scaling up the number of new BCR facilities to over 1000 per year by 2030 becomes imperative. The scale-up of BCR facilities, however, brings challenges in terms of operation modes.
Most prominent the emission control of the facilities needs to be under control to be possible to operate the machine under the given emission limits.

In light of these challenges, this paper seeks to introduce an approach to predict the NOx emissions of BCR plants, which can be seen as the most critical emission limit.
Specifically, our focus lies in demonstrating how machine learning algorithms can be harnessed to model the intricate states of highly complex pyrolysis machines, thereby enabling real-time prediction of production values based on the current operational state.

With this work, we relate to multiple strands of literature.
Several papers use machine learning to model biochar-related processes already, but is primarily interested in yield prediction \citep{LI2022127511, HAI2023103071}.
Similarly, in domains other than biochar, machine learning has successfully been used to predict NOx emissions during physical and chemical processes, but remains focused on coal-based power plants \citep{MATSUZAKI20232858,WU2023127044}, diesel engines \citep{samosirNOxEmissionsPrediction2024} and ammonia-hydrogen combustion \citep{CHATURVEDI2023101406} so far.

However, for predicting NOx emissions of biochar plants there is no comparable work yet.
Beyond this research gap, we also go a step further by subsequently using the predictions as \textit{surrogates} during an optimization procedure.
While we focus on the prediction of NOx emissions here, the machine learning approach can be easily expanded to predict multiple outcomes of the biochar process, and potentially adapted to (physical and chemical) processes other than biochar production.
Overall, we thus believe we make a methodologically novel contribution to a timely and pressing issue.

\section{Reactors: PRYEG and ARTi}
Our approach was employed on two pyrolysis machines, manufactured by PYREG GmbH~\citep{pyreg} and ARTi \citep{arti}, exemplarily illustrated in Figure~\ref{fig:pyreg}.
These machines are equipped with a range of sensors (see sensor symbol in the figure), including those for mass flow, moisture content, and temperature, among others.
Moreover, they offer diverse control settings (see screw symbol in the figure), such as valve positions, ventilator adjustments, and oxygen flow control. 
It's important to note that in Figure~\ref{fig:pyreg}, only select components of this machine are depicted.
In the following, we refer to the control settings as ``machine states'', which serve as the input features for the model, while the sensors represent the target values predicted by the model.

The operational process involves introducing various feedstocks (e.g. sewage sludge) from the left side of the machine, which are subsequently conveyed to the reactor.
Within this reactor, the feedstock undergoes controlled pyrolysis under specific temperatures (from \SI{600}{} to \SI{1200}{\celsius}) and different duration times.
The resultant biochar is collected, and its moisture content is assessed as part of the process.


Contrary to what is depicted in the picture, most pyrolysis machines on the market do not continuously measure NOx values.
In the case of the PYREG machine, NOx data was collected over two months of operation, whereas for the ARTi machine, NOx measurements were only conducted over two days using an external device.
Consequently, predicting NOx values based solely on machine states offers a viable solution for continuous emission monitoring.


\section{Machine Learning Approach}


\begin{figure}[ht]
	\centering
	\begin{subfigure}{0.5\linewidth}
		\includegraphics[width = \linewidth]{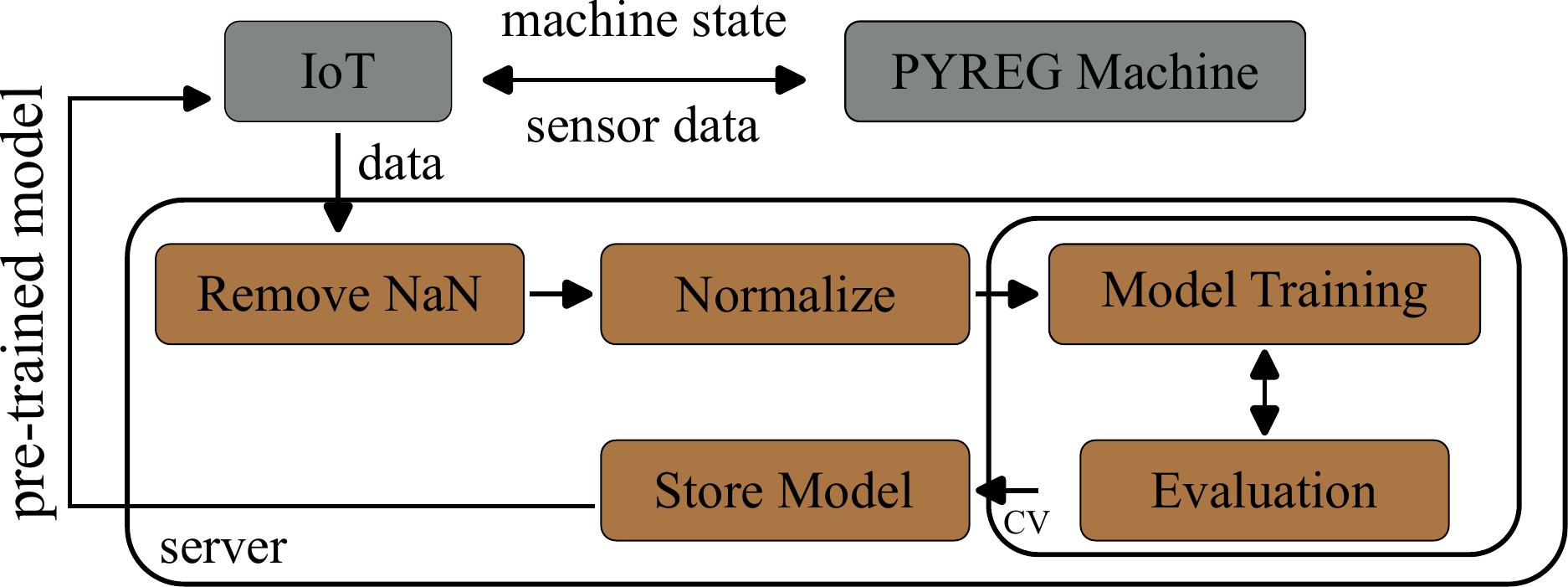}
            \caption{} 
            \label{fig:pipeline}
	\end{subfigure}
	\begin{subfigure}{0.3\linewidth}
            \includegraphics[width=\linewidth]{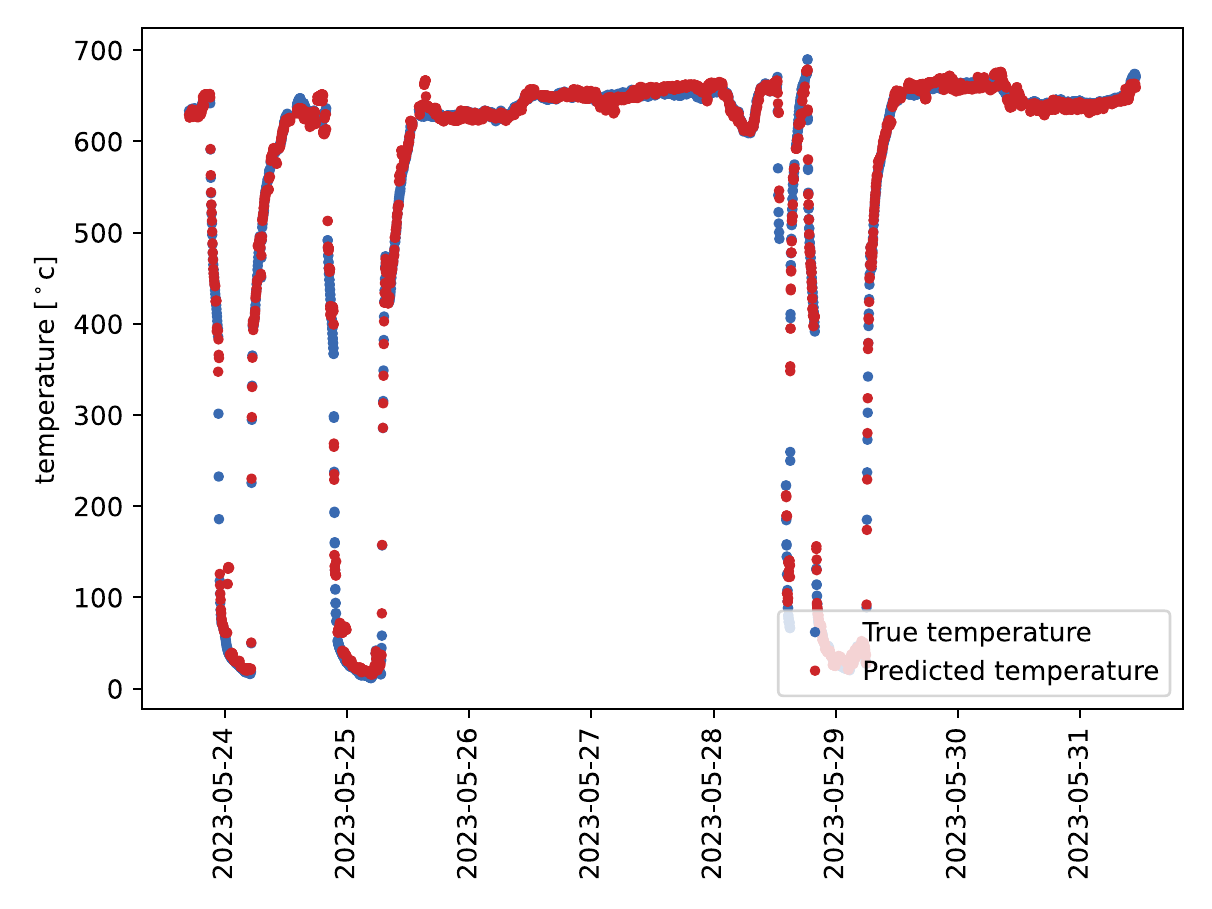}
		\caption{}
		\label{fig:subfigA}
         \end{subfigure}
         \caption{Figure \ref{fig:pipeline}: Overview of the used machine learning pipeline. The grey boxes represent the parts that are employed at the machine, while the brown boxes show the pipeline to pre-train the model on a server. Figure \ref{fig:subfigA}: First test of predicting the reactor temperature which is a key feature of how much NOx is produced.}
	\label{fig:pipeline2}
\end{figure}

To mitigate NOx emissions in biochar production, our initial strategy involved replicating the feedstock-to-biochar conversion process.
This was accomplished by leveraging sensor data and machine states to construct a digital twin of the pyrolysis machine.

For this purpose, we employed a Random Forest Regressor~\citep{random} to predict various sensor outputs (target values) based on the machine states (input features).
Notably, the choice of Random Forest was influenced by its documented success in optimizing operational parameters in industrial settings~\citep{alshraideh2020process} and its proficiency in controlling and monitoring industrial machinery~\citep{8709808}.

Figure~\ref{fig:pipeline} shows our training and evaluation pipeline.
At its core is an IoT device integrated with the machine, capable of altering machine states and capturing sensor data.
This data, encompassing both machine states and sensor readings, is transmitted to a server.
The components installed on the machine are indicated by grey boxes, while the brown boxes depict the server-based pre-training pipeline.
We orchestrated the entire process using Kedro~\citep{Alam_Kedro_2023} and scikit-learn~\citep{pedregosa2011scikit}.

Initially, a model pre-trained on the server utilized two years of historical machine operation data.
This data was structured into a flat table format suitable for Random Forest processing~\citep{time-forest}.
We conducted data cleaning by removing missing values and normalizing the dataset.
Then, a five-fold cross-validation (CV) was executed to determine optimal hyperparameters.
This phase involved evaluating various hyperparameters, selecting the combination that best predicted sensor data from machine states.
The hyperparameter tuning focused on the mean squared error, exploring combinations of a number of estimators (10-100), min\_samples\_split (2-5) and min\_samples\_leaf (1-3).
The final model was then stored for subsequent deployment to the IoT device.

Post pre-training, the model was transferred to the IoT device.
Due to the device's limited resources, we used a simplified Random Forest model (maximum tree depth of 2).
In future iterations, we aim to upgrade to a more robust computing solution.

To evaluate our approach, we conducted an initial testing phase using the PYREG reactor. During this two-week period of continuous operation, various operational modes were tested to assess performance under different conditions. The pre-trained model was retrained on the IoT device every two hours using the most recent data, striking a balance between maintaining strong predictive performance and not overloading the device's processing capabilities.

In this initial test phase, our focus was on predicting machine states; for example, Figure~\ref{fig:subfigA} illustrates the predicted reactor temperatures alongside the actual measured values. Since continuous NOx measurements are not always available for these machines, we first evaluated our model using consistently available sensor data.

\section{NOx Evaluation \& Optimization}


\begin{figure}[ht]
	\centering
	\begin{subfigure}{0.32\linewidth}
		\includegraphics[width=\linewidth]{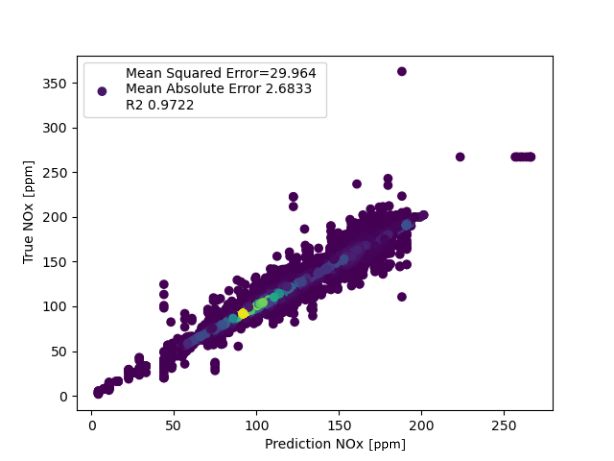}
            \caption{} 
            \label{fig:subfigB}
	\end{subfigure}
 	\begin{subfigure}{0.32\linewidth}
		\includegraphics[width=\linewidth]{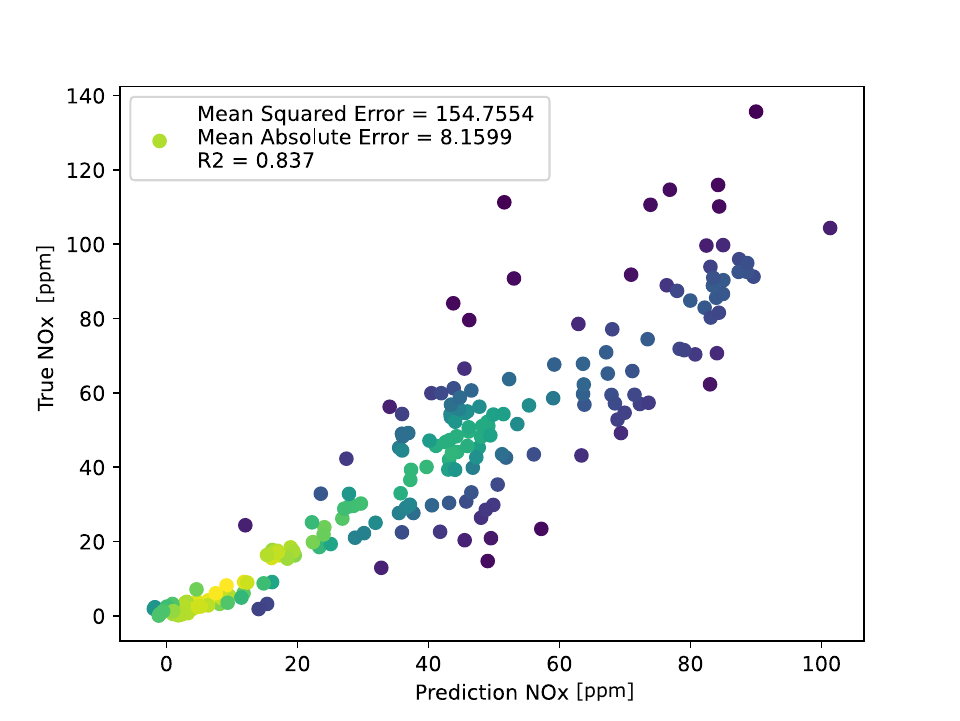}
            \caption{} 
            \label{fig:subfigD}
	\end{subfigure}
 	\begin{subfigure}{0.32\linewidth}
		\includegraphics[width=\linewidth]{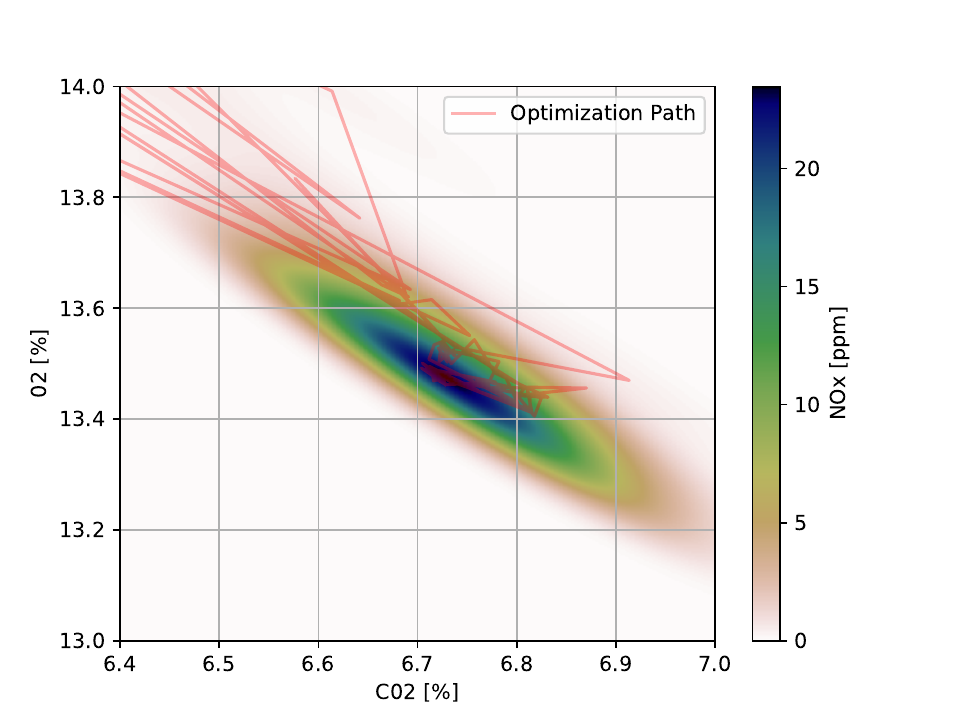}
            \caption{} 
            \label{fig:opti}
	\end{subfigure}
         \caption{Figure \ref{fig:subfigB} and \ref{fig:subfigD}: Show the prediction of NOx value using historical data from the PYREG and the ARTi reactor respectively. Figure \ref{fig:opti} shows the optimization path for the ARTi data to minimize NOx and have CO2 and O2 (as two example values) constrained between CO2 0-10 \% and O2 0-20 \%.
         }
	\label{fig:pipeline3}
\end{figure}

In the second testing phase, we focused on the primary objective of the study: predicting the NOx values for the two machines.
Figure~\ref{fig:subfigB} presents the NOx predictions based on two months of continuous sensor data from the PYREG machine.
This prediction was achieved using the same pipeline, which was fine-tuned with additional machine states such as temperature data (see Figure~\ref{fig:subfigA}).
Similarly, Figure~\ref{fig:subfigD} illustrates the NOx predictions based on two days of external NOx measurements performed on one of the ARTi reactors.

Both tests demonstrated strong predictive performance, with an R2 score of 0.97 for the PYREG reactor and 0.84 for the ARTi reactor.
The lower R2 score for the ARTi reactor is primarily attributed to the limited amount of data available compared to the PYREG case.


Figure~\ref{fig:opti} highlights the relationship between O2 and CO2 concentration during the minimization of NOx based on the ARTi data.
The optimization was done by constraining the prediction of O2 and CO2 of the Random Forest Regressor should be constrained by 0\% < O2 < 10 \% and 0\% < O2 < 20 \%.
All constrains can be set by the user to obtain an optimization based on its needs.

\section{Conclusion \& Future Work}
In this paper, we present a proof-of-concept approach for modeling the state of a pyrolysis machine and predicting NOx values using machine learning, thereby enabling machine monitoring without the need for continuous NOx sensors.
We demonstrated the model's performance on two different pyrolysis machines from distinct manufacturers, with both tests yielding promising results for continuous NOx prediction.

By extending this approach to predict other metrics, such as throughput or biochar yield, it is possible to develop an optimization model that simultaneously minimizes NOx emissions while maximizing biochar production.

\end{document}

%% file: math_commands.tex
\usepackage{amsmath,amsfonts,bm}

\def\eqref#1{equation~\ref{#1}}

\def\1{\bm{1}}

\DeclareMathAlphabet{\mathsfit}{\encodingdefault}{\sfdefault}{m}{sl}
\SetMathAlphabet{\mathsfit}{bold}{\encodingdefault}{\sfdefault}{bx}{n}













%% file: main.bbl
\begin{thebibliography}{20}
\providecommand{\natexlab}[1]{#1}
\providecommand{\url}[1]{\texttt{#1}}
\expandafter\ifx\csname urlstyle\endcsname\relax
  \providecommand{\doi}[1]{doi: #1}\else
  \providecommand{\doi}{doi: \begingroup \urlstyle{rm}\Url}\fi

\bibitem[Alam et~al.(2023)]{Alam_Kedro_2023}
Sajid Alam et~al.
\newblock {Kedro}, July 2023.
\newblock URL \url{https://github.com/kedro-org/kedro}.

\bibitem[Alshraideh et~al.(2020)Alshraideh, Del~Castillo, and
  Del~Val]{alshraideh2020process}
Hussam Alshraideh, Enrique Del~Castillo, and Alain~Gil Del~Val.
\newblock Process control via random forest classification of profile signals:
  An application to a tapping process.
\newblock \emph{Journal of Manufacturing Processes}, 2020.

\bibitem[Anderson \& Peters(2016)Anderson and Peters]{negemissions}
Kevin Anderson and Glen Peters.
\newblock The trouble with negative emissions.
\newblock \emph{Science}, 2016.
\newblock \doi{10.1126/science.aah4567}.

\bibitem[ARTi(2024)]{arti}
ARTi.
\newblock {ARTi}, 2024.
\newblock Online: \url{www.arti.com/}.

\bibitem[Board et~al.(2019)Board, National Academies~of Sciences, Medicine,
  et~al.]{board2019negative}
Ocean~Studies Board, Engineering National Academies~of Sciences, Medicine,
  et~al.
\newblock Negative emissions technologies and reliable sequestration: a
  research agenda.
\newblock 2019.

\bibitem[Brownlee(2020)]{time-forest}
Jason Brownlee.
\newblock Time series, 2020.
\newblock Online:
  \url{https://machinelearningmastery.com/random-forest-for-time-series-forecasting/}.

\bibitem[Chai \& Zhao(2020)Chai and Zhao]{8709808}
Zheng Chai and Chunhui Zhao.
\newblock Enhanced random forest with concurrent analysis of static and dynamic
  nodes for industrial fault classification.
\newblock \emph{IEEE Transactions on Industrial Informatics}, 2020.
\newblock \doi{10.1109/TII.2019.2915559}.

\bibitem[Chaturvedi et~al.(2023)Chaturvedi, Santhosh, Mashruk, Yadav, and
  Valera-Medina]{CHATURVEDI2023101406}
Shivansh Chaturvedi, R.~Santhosh, Syed Mashruk, Rajneesh Yadav, and Agustin
  Valera-Medina.
\newblock Prediction of nox emissions and pathways in premixed
  ammonia-hydrogen-air combustion using cfd-crn methodology.
\newblock \emph{Journal of the Energy Institute}, 111:\penalty0 101406, 2023.
\newblock ISSN 1743-9671.
\newblock \doi{https://doi.org/10.1016/j.joei.2023.101406}.
\newblock URL
  \url{https://www.sciencedirect.com/science/article/pii/S1743967123002350}.

\bibitem[Hai et~al.(2023)Hai, Bharath, Patah, Daud, K., Show, and
  Banat]{HAI2023103071}
Abdul Hai, G.~Bharath, Muhamad Fazly~Abdul Patah, Wan Mohd Ashri~Wan Daud,
  Rambabu K., PauLoke Show, and Fawzi Banat.
\newblock Machine learning models for the prediction of total yield and
  specific surface area of biochar derived from agricultural biomass by
  pyrolysis.
\newblock \emph{Environmental Technology \& Innovation}, 30:\penalty0 103071,
  2023.
\newblock ISSN 2352-1864.
\newblock \doi{https://doi.org/10.1016/j.eti.2023.103071}.
\newblock URL
  \url{https://www.sciencedirect.com/science/article/pii/S2352186423000676}.

\bibitem[Ho(1995)]{random}
Tin~Kam Ho.
\newblock Random decision forests.
\newblock In \emph{Proceedings of 3rd international conference on document
  analysis and recognition}, volume~1, pp.\  278--282. IEEE, 1995.

\bibitem[Initiative(2023)]{ibi}
European Biochar IndustryInternational~Biochar Initiative.
\newblock Global biochar market report, 2023.
\newblock Online: \url{biochar-international.org}.

\bibitem[Li et~al.(2022)Li, Gupta, and You]{LI2022127511}
Yize Li, Rohit Gupta, and Siming You.
\newblock Machine learning assisted prediction of biochar yield and composition
  via pyrolysis of biomass.
\newblock \emph{Bioresource Technology}, 359:\penalty0 127511, 2022.
\newblock ISSN 0960-8524.
\newblock \doi{https://doi.org/10.1016/j.biortech.2022.127511}.
\newblock URL
  \url{https://www.sciencedirect.com/science/article/pii/S0960852422008409}.

\bibitem[Matsuzaki et~al.(2023)Matsuzaki, Kiribuchi, and
  Shimizu]{MATSUZAKI20232858}
A.~Matsuzaki, D.~Kiribuchi, and K.~Shimizu.
\newblock Machine learning approach to nox prediction for scr process of
  thermal power plant.
\newblock \emph{IFAC-PapersOnLine}, 56\penalty0 (2):\penalty0 2858--2864, 2023.
\newblock ISSN 2405-8963.
\newblock \doi{https://doi.org/10.1016/j.ifacol.2023.10.1401}.
\newblock URL
  \url{https://www.sciencedirect.com/science/article/pii/S2405896323018098}.
\newblock 22nd IFAC World Congress.

\bibitem[Pedregosa et~al.(2011)]{pedregosa2011scikit}
Fabian Pedregosa et~al.
\newblock Scikit-learn: Machine learning in python.
\newblock \emph{Journal of machine learning research}, 12\penalty0
  (Oct):\penalty0 2825--2830, 2011.

\bibitem[PYREG(2024)]{pyreg}
PYREG.
\newblock {PYREG GmbH}, 2024.
\newblock Online: \url{pyreg.com}.

\bibitem[Rogelj et~al.(2018)]{rogelj2018scenarios}
Joeri Rogelj et~al.
\newblock Scenarios towards limiting global mean temperature increase below 1.5
  c.
\newblock \emph{Nature Climate Change}, 2018.

\bibitem[Samosir et~al.(2024)Samosir, Quach, Chul, and
  Lim]{samosirNOxEmissionsPrediction2024}
Bernike~Febriana Samosir, Nhu~Y. Quach, Oh~Kwang Chul, and Ocktaeck Lim.
\newblock {NOx} emissions prediction in diesel engines: a deep neural network
  approach.
\newblock \emph{Environmental Science and Pollution Research}, 31\penalty0
  (1):\penalty0 713--722, January 2024.
\newblock ISSN 1614-7499.
\newblock \doi{10.1007/s11356-023-30937-3}.
\newblock URL \url{https://doi.org/10.1007/s11356-023-30937-3}.

\bibitem[Shukla et~al.(2022)]{ipcc}
P.R. Shukla et~al.
\newblock Climate change 2022 - mitigation of climate change: Working group iii
  contribution to the sixth assessment report of the intergovernmental panel on
  climate change.
\newblock 2022.
\newblock \doi{10.1017/9781009157926}.

\bibitem[Smith et~al.(2016)]{smith2016biophysical}
Pete Smith et~al.
\newblock {Biophysical and economic limits to negative CO2 emissions}.
\newblock \emph{Nature climate change}, 2016.

\bibitem[Wu et~al.(2023)Wu, Zhang, and Dong]{WU2023127044}
Zheng Wu, Yue Zhang, and Ze~Dong.
\newblock Prediction of nox emission concentration from coal-fired power plant
  based on joint knowledge and data driven.
\newblock \emph{Energy}, 271:\penalty0 127044, 2023.
\newblock ISSN 0360-5442.
\newblock \doi{https://doi.org/10.1016/j.energy.2023.127044}.
\newblock URL
  \url{https://www.sciencedirect.com/science/article/pii/S0360544223004383}.

\end{thebibliography}
